\documentclass[conference]{IEEEtran}
\IEEEoverridecommandlockouts
\usepackage{cite}
\usepackage{url}
\makeatletter
\newcommand*{\rom}[1]{\expandafter\@slowromancap\romannumeral #1@}
\makeatother
\usepackage{amsmath,amssymb,amsfonts}
\usepackage{algorithmic}
\usepackage{graphicx}
\usepackage{textcomp}
\usepackage{xcolor}
\def\BibTeX{{\rm B\kern-.05em{\sc i\kern-.025em b}\kern-.08em
    T\kern-.1667em\lower.7ex\hbox{E}\kern-.125emX}}
\begin{document}

\title{Spoken Speech Enhancement using EEG\\
{
}
}

\author{\IEEEauthorblockN{Gautam Krishna}
\IEEEauthorblockA{\textit{Brain Machine Interface Lab} \\
\textit{The University of Texas at Austin}\\
Austin, Texas \\
}
\and
\IEEEauthorblockN{Co Tran}
\IEEEauthorblockA{\textit{Brain Machine Interface Lab} \\
\textit{The University of Texas at Austin}\\
Austin, Texas \\
}
\and
\IEEEauthorblockN{Yan Han}
\IEEEauthorblockA{\textit{Brain Machine Interface Lab} \\
\textit{The University of Texas at Austin}\\
Austin, Texas \\
}
\and
\IEEEauthorblockN{Mason Carnahan}
\IEEEauthorblockA{\textit{Brain Machine Interface Lab} \\
\textit{The University of Texas at Austin}\\
Austin, Texas \\
}
\and
\IEEEauthorblockN{Ahmed H Tewfik}
\IEEEauthorblockA{\textit{Brain Machine Interface Lab} \\
\textit{The University of Texas at Austin}\\
Austin, Texas  \\
}
}

\maketitle

\begin{abstract}
In this paper we demonstrate spoken speech enhancement using electroencephalography (EEG) signals using a generative adversarial network (GAN) based model, gated recurrent unit (GRU) regression based model, temporal convolutional network (TCN) regression model and finally using a mixed TCN GRU regression model.

We compare our EEG based speech enhancement results with traditional log minimum mean-square error (MMSE) speech enhancement algorithm and our proposed methods demonstrate significant improvement in speech enhancement quality compared to the traditional method. 
Our overall results demonstrate that EEG features can be used to clean speech recorded in presence of background noise. 
To the best of our knowledge this is the first time a spoken speech enhancement is demonstrated using EEG features recorded in parallel with spoken speech.
\end{abstract}


\section{Introduction}
Speech enhancement is the process of improving the quality of speech whose quality was degraded due to additive noise. Speech enhancement is a critical preprocessing method used to improve the performance of automatic speech recognition (ASR) systems operating in presence of background noise. Noisy speech is first fed into a speech enhancement system to produce enhanced speech which is then fed into the ASR model. Speech enhancement systems also plays critical role in improving the quality of speech used in devices like hearing aids and cochlear implants. 

In references \cite{berouti1979enhancement,ephraim1992statistical} authors demonstrated speech enhancement using classical methods. Recently researchers have started applying deep learning methods for performing speech enhancement as indicated in the following references \cite{parveen2004speech,weninger2015speech}. In references \cite{pascual2017segan} authors demonstrated speech enhancement using generative adversarial networks (GAN)\cite{goodfellow2014generative}. 

Electroencephalography (EEG) is a non invasive way of measuring electrical activity of human brain. In \cite{krishna2019speech} authors demonstrated that EEG features can be used to overcome the performance loss of ASR systems in presence of background noise. Though references \cite{krishna20,krishna2019state,krishna2019speech} demonstrated isolated and continuous speech recognition using EEG signals for various experimental conditions, they didn't specifically study the speech enhancement problem. In this paper we demonstrate that EEG features can be used to improve the quality of speech recorded in presence of background noise. We make use of GAN, gated recurrent unit (GRU) \cite{chung2014empirical}, temporal convolutional (TCN) \cite{bai2018empirical} networks to demonstrate speech enhancement using EEG features. We further compare our obtained results with traditional log minimum mean-square error (MMSE) speech enhancement algorithm and our results demonstrate significant improvement in speech enhancement quality compared to the traditional method. 

In \cite{das2017eeg} authors demonstrated EEG based attention driven speech enhancement using wiener filters where EEG was used to detect auditory attention where as in this paper we demonstrate speech enhancement for "Spoken" speech using EEG features and auditory attention detection module is not required for performing speech enhancement. Our idea is mainly inspired by the results demonstrated in \cite{krishna2019speech} where authors demonstrated EEG features are less affected by external background noise. To the best of our knowledge this is the first time a spoken speech enhancement is demonstrated using EEG features recorded in parallel with spoken speech.

\section{Design of Experiments for building Training and Test Set}
 
We used Data set A used by authors in \cite{krishna20} as training set. The Data set A consists of simultaneous speech and EEG recordings from 10 subjects. This data was recorded in absence of externally created background noise but a background noise of 40 dB due to the sound of lab ventilation fan was observed. For the sake of the simplicity of the study we would neglect this 40 dB noise effect and would consider training data set as clean. 

We used Data set B used by authors in \cite{krishna20} as test set. The Data set B consists of simultaneous speech and EEG recordings from 8 subjects recorded in presence of external background noise of 65 dB. For collecting data for training and test set, among the total number of subjects, five subjects took part in both the experiments for Data sets. 

We used Brain product's ActiChamp EEG recording hardware. Our EEG cap had 32 wet EEG electrodes including one electrode as ground as shown in Figure 1. We used EEGLab to obtain the EEG sensor location mapping. It is based on standard 10-20 EEG sensor placement method for 32 electrodes. 

\begin{figure}[h]
\begin{center}
\includegraphics[height=3cm,width=0.25\textwidth,trim={1cm 1cm 1cm 0.1cm},clip]{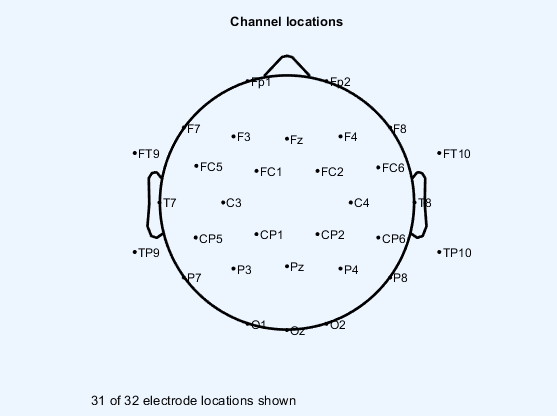}
\caption{EEG channel locations for the cap used in our experiments} 
\label{1vsall}
\end{center}
\end{figure}

\section{EEG and Speech feature extraction details}

We followed the same methodology used by authors in references \cite{krishna20,krishna2019speech} for EEG and speech preprocessing. EEG signals were sampled at 1000Hz and a fourth order IIR band pass filter with cut off frequencies 0.1Hz and 70Hz was applied. A notch filter with cut off frequency 60 Hz was used to remove the power line noise.
EEGlab's Independent component analysis (ICA) toolbox was used to remove other biological signal artifacts like electrocardiography (ECG), electromyography (EMG), electrooculography (EOG) etc from the EEG signals. We extracted five statistical features for EEG, namely root mean square, zero crossing rate,moving window average,kurtosis and power spectral entropy \cite{krishna2019speech,krishna20,krishna200speech}. So in total we extracted 31(channels) X 5 or 155 features for EEG signals. The EEG features were extracted at a sampling frequency of 100Hz for each EEG channel.  

The recorded speech signal was sampled at 16KHz frequency. We extracted Mel-frequency cepstrum coefficients (MFCC) as features for speech signal.
We extracted MFCC 13 features and the MFCC features were also sampled at 100Hz same as the sampling frequency of EEG features

\begin{figure}[h]
\centering
\includegraphics[height=5cm, width=0.4
\textwidth,trim={0.1cm 0.1cm 0.1cm 0.1cm},clip]{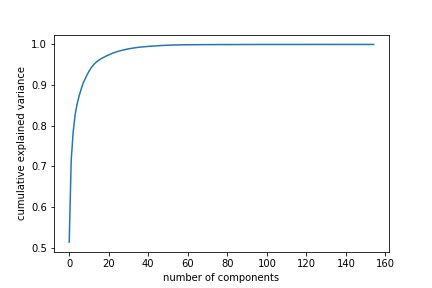}
\caption{Explained variance plot}
\label{1vsall}
\end{figure}

\section{EEG Feature Dimension Reduction Algorithm Details}

We reduced the 155 EEG features to a dimension of 30 by applying Kernel Principle Component Analysis (KPCA) to denoise the EEG feature space as demonstrated by authors in \cite{krishna20,krishna2019state}. We plotted cumulative explained variance versus number of components to identify the right feature dimension as shown in Figure 2. We used KPCA with polynomial kernel of degree 3 \cite{krishna2019speech,krishna20,krishna200speech}.

\section{Speech Enhancement models}

We used four different types of model for performing speech enhancement using EEG features. We first performed experiments using a simple gated recurrent units (GRU) \cite{chung2014empirical} regression model followed by speech enhancement experiments using a generative adversarial networks (GAN) \cite{goodfellow2014generative} model, followed by speech enhancement experiments using temporal convolutional network (TCN) \cite{bai2018empirical} regression model and finally we performed speech enhancement using a mixture of TCN, GRU regression model. 

In the below sub sections we explain the architecture of our models and experiment set up details. Our GAN model architecture is different from the ones used by authors in references \cite{pascual2017segan}. We added Gaussian noise with zero mean and standard deviation 10 to the recorded MFCC features from training set to generate noisy MFCC features.
These noisy MFCC features will be used during training of the models as explained in below sub sections. The gaussian noise was not added to the EEG features from training set as our hypothesis was effect of background noise on EEG features is negligible \cite{krishna2019speech}. The gaussian noise was not added to the test set data as it was already collected in presence of externally created background noise.

\subsection{GRU Regression Model}

Our GRU regression model consists of two layers of GRU with 128 hidden units in first layer and with 64 hidden units in second layer followed by a time distributed dense layer with 13 hidden units. The GRU regression model architecture is shown in Figure 3. The model was trained for 1000 epochs to observe loss convergence and adam optimizer was used. The Batch size was set to 200. Mean squared error (MSE) was used as the loss function. The validation split was set to 0.1. The Figure 4 shows training and validation loss convergence of the model. 

During training time, we concatenate the generated noisy MFCC features (after adding gaussian noise) and recorded EEG features from the training set and feed it as a single vector input to the GRU regression model and corresponding clean MFCC features from training set of dimension 13 are set as targets.  

During test time, we concatenate the MFCC and EEG features from test set and feed it as a single vector input to the trained GRU regression model to output corresponding enhanced MFCC. Griffin Lim reconstruction \cite{griffin1984signal} algorithm is used to convert enhanced MFCC to speech. 

\begin{figure}[h]
\centering
\includegraphics[height=5cm, width=0.5
\textwidth,trim={0.1cm 0.1cm 0.1cm 0.1cm},clip]{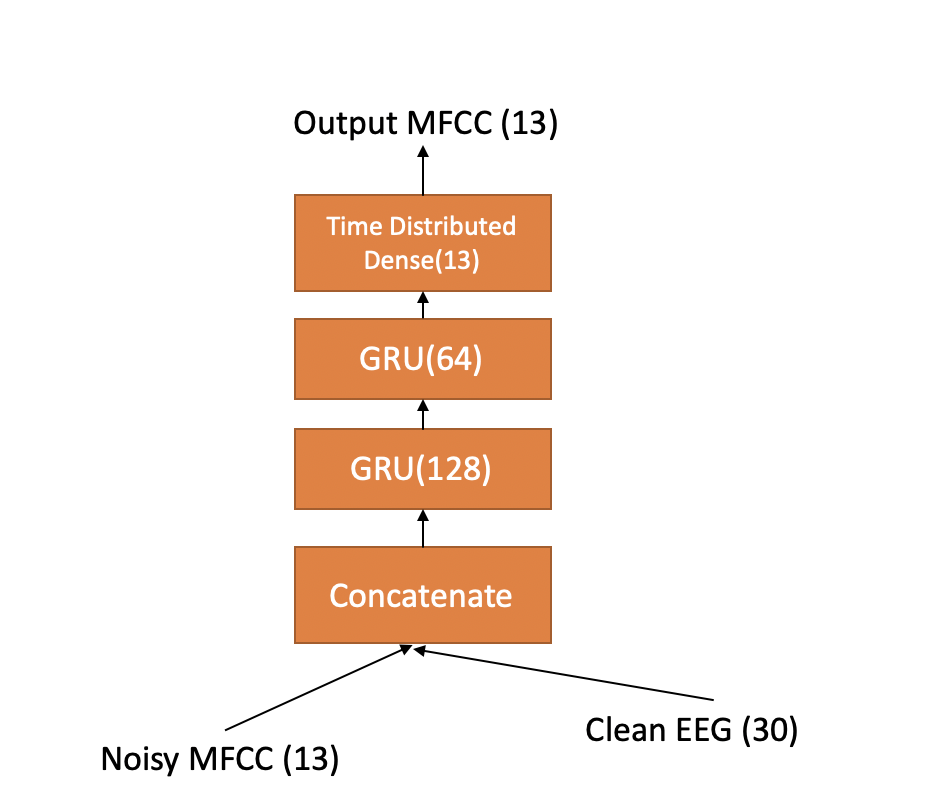}
\caption{GRU regression model}
\label{1vsall}
\end{figure}

\begin{figure}[h]
\centering
\includegraphics[height=5cm, width=0.5
\textwidth,trim={0.1cm 0.1cm 0.1cm 0.1cm},clip]{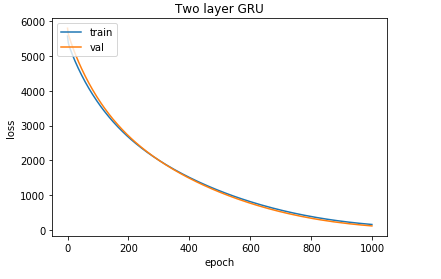}
\caption{GRU regression model training loss convergence}
\label{1vsall}
\end{figure}

\subsection{TCN Regression Model}

Our TCN regression model consists of a single layer of TCN with 128 filters followed by a time distributed dense layer with 13 hidden units. The TCN regression model architecture is shown in Figure 5. The model was trained for 1000 epochs to observe loss convergence and adam optimizer \cite{kingma2014adam} was used. The Batch size was set to 200. Mean squared error (MSE) was used as the loss function. The validation split was set to 0.1. 

\begin{figure}[h]
\centering
\includegraphics[height=5cm, width=0.5
\textwidth,trim={0.1cm 0.1cm 0.1cm 0.1cm},clip]{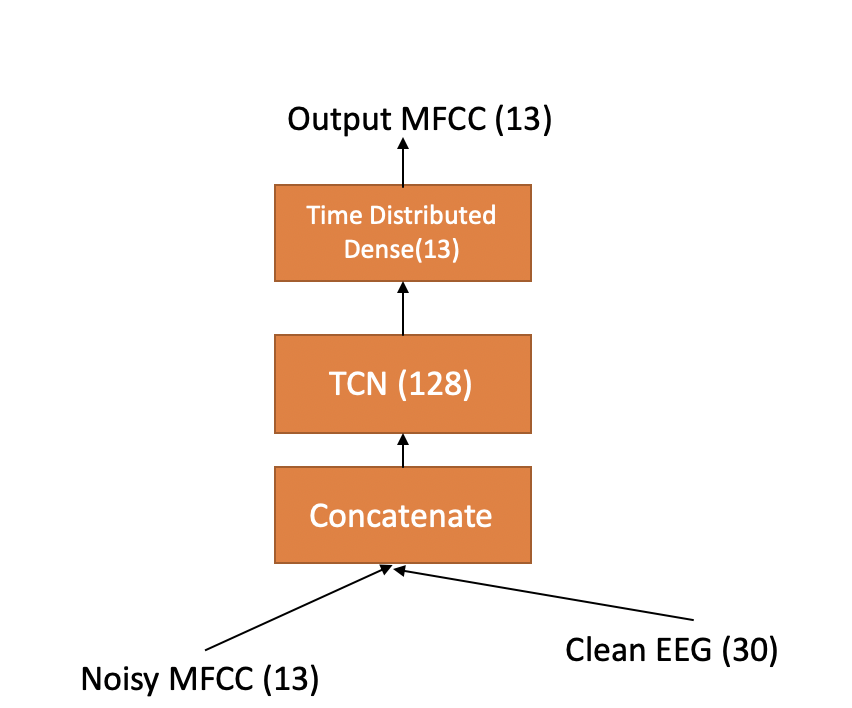}
\caption{TCN regression model}
\label{1vsall}
\end{figure}

\subsection{GRU-TCN Regression Model}

Our GRU -TCN regression model consists of two layers of GRU with 128 hidden units in first layer and with 64 hidden units in second layer followed by a  single layer of TCN with 32 filters followed by a time distributed dense layer with 13 hidden units. A dropout regularization \cite{srivastava2014dropout} with dropout rate 0.2 is applied between the TCN and GRU layer. 
The GRU-TCN regression model architecture is shown in Figure 6. The model was trained for 1000 epochs to observe loss convergence and adam optimizer \cite{kingma2014adam} was used. The Batch size was set to 200. Mean squared error (MSE) was used as the loss function. The validation split was set to 0.1.  The Figure 7 shows training and validation loss convergence of the model. 

\begin{figure}[h]
\centering
\includegraphics[height=5cm, width=0.5
\textwidth,trim={0.1cm 0.1cm 0.1cm 0.1cm},clip]{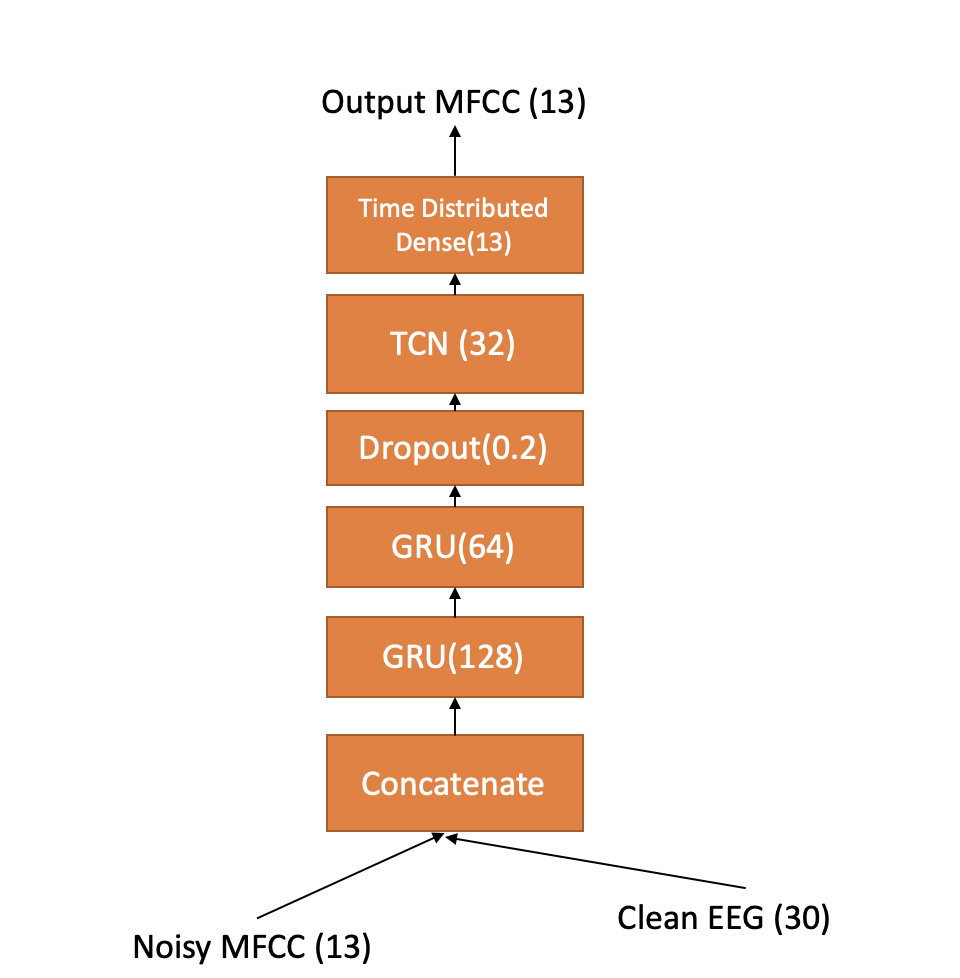}
\caption{GRU-TCN regression model}
\label{1vsall}
\end{figure}

\begin{figure}[h]
\centering
\includegraphics[height=5cm, width=0.5
\textwidth,trim={0.1cm 0.1cm 0.1cm 0.1cm},clip]{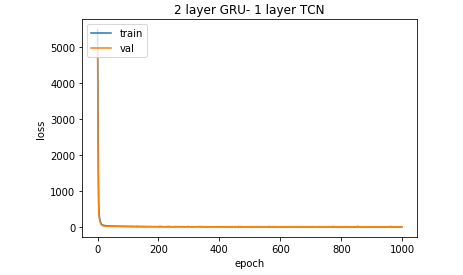}
\caption{GRU-TCN regression model training loss convergence}
\label{1vsall}
\end{figure}

\subsection{GAN Model}

Generative Adversarial Network (GAN) consists of two networks namely the generator model and the discriminator model which are trained simultaneously. The generator model learns to generate data from a latent space and the discriminator model evaluates whether the data generated by the generator is fake or is from true data distribution. The training objective of the generator is to fool the discriminator. 

Our main motivation behind using GAN model was in the case of GAN, the loss function is learned during training of the model instead of using a fixed loss function in the case of GRU or TCN or GRU-TCN regression model ( ie: MSE). 

Our generator model consists of two parallel GRU's with 128 and 64 hidden units in each layer. The outputs of the two parallel GRU's are concatenated and fed into TCN layer with 32 filters followed by a time distributed dense layer of 13 hidden units. The architecture of discriminator model is similar to that of the generator model but instead of the time distributed dense layer, a dense layer with single hidden unit sigmoid activation is used. The last time step output of the preceding TCN layer is fed into the dense layer. 

During training time, the generator always takes noisy MFCC ( obtained after adding gaussian noise to clean MFCC from training set) and clean EEG ( from training set) as input pairs and outputs fake MFCC. The Generator model architecture is shown in Figure 8.
The discriminator can take three possible pairs of inputs during training. Let $P_{f}$ be the sigmoid output of the discriminator for (fake MFCC, clean EEG) pair input, $P_{c}$ be the sigmoid output of the discriminator for (clean MFCC, clean EEG) pair input and $P_{n}$ be the sigmoid output of the discriminator for
(noisy MFCC, clean EEG) pair input, then we can define the loss function of the generator as $-\log (P_{f})$ and loss function of the discriminator as $-\log (1-P_{f}) - \log (1-P_{n}) - \log (P_{c})$ for speech enhancement. The model was trained for 100 epochs using adam optimizer with a batch size of 32. The Discriminator model architecture is shown in Figure 9. Input 1, Input 2 in the figure refers to the three possible pairs of input for the discriminator during training. Figures 10 and 11 shows the training loss for the generator and discriminator models. 

During test time, the trained generator model takes (MFCC, EEG) input pair from the test set and outputs enhanced MFCC and we use griffin lim reconstruction algorithm to convert enhanced MFCC to speech.

\begin{figure}[h]
\centering
\includegraphics[height=5cm, width=0.5
\textwidth,trim={0.1cm 0.1cm 0.1cm 0.1cm},clip]{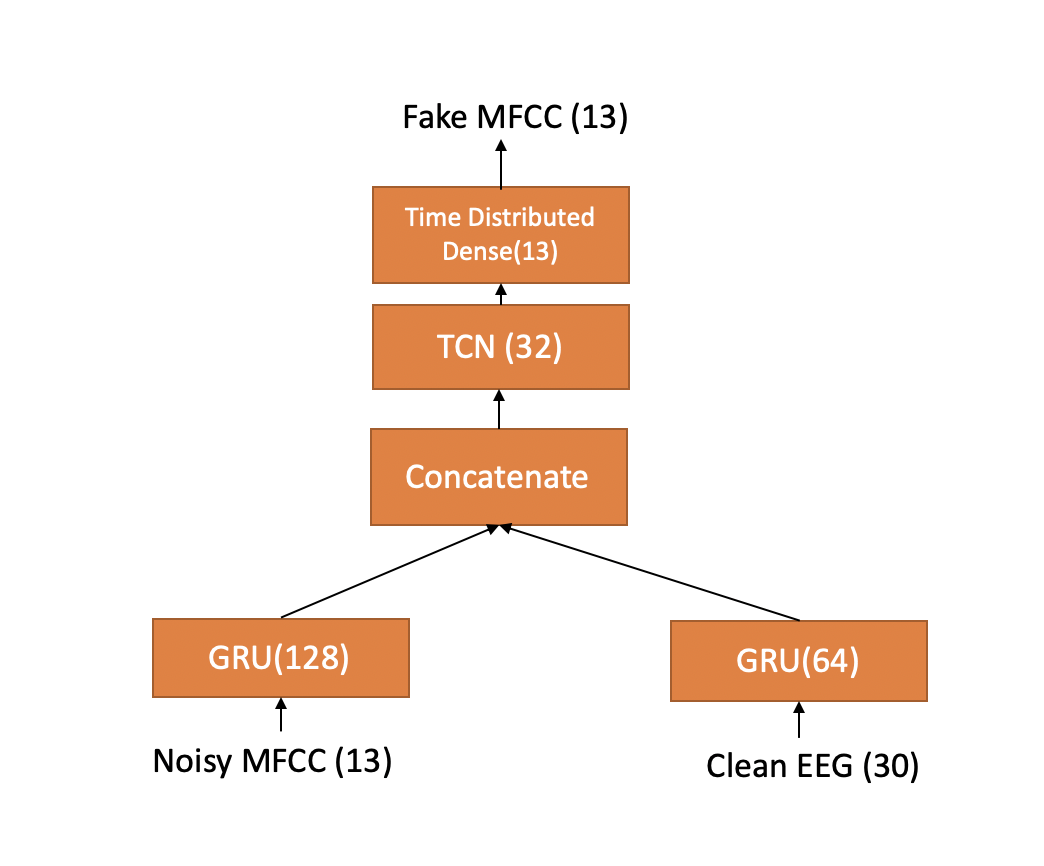}
\caption{Generator in GAN model}
\label{1vsall}
\end{figure}

\begin{figure}[h]
\centering
\includegraphics[height=5cm, width=0.5
\textwidth,trim={0.1cm 0.1cm 0.1cm 0.1cm},clip]{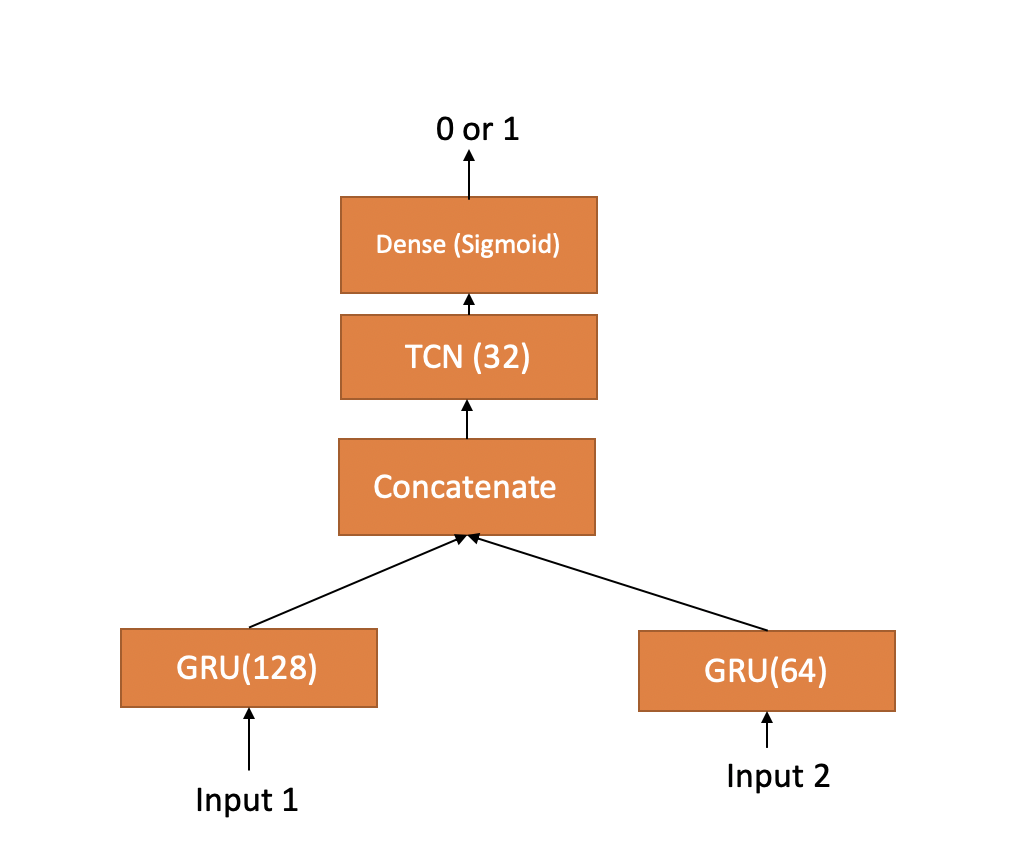}
\caption{Discriminator in GAN model}
\label{1vsall}
\end{figure}

\begin{figure}[h]
\centering
\includegraphics[height=5cm, width=0.5
\textwidth,trim={0.1cm 0.1cm 0.1cm 0.1cm},clip]{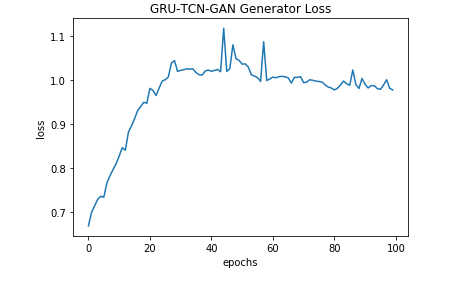}
\caption{Generator training loss convergence}
\label{1vsall}
\end{figure}

\begin{figure}[h]
\centering
\includegraphics[height=5cm, width=0.5
\textwidth,trim={0.1cm 0.1cm 0.1cm 0.1cm},clip]{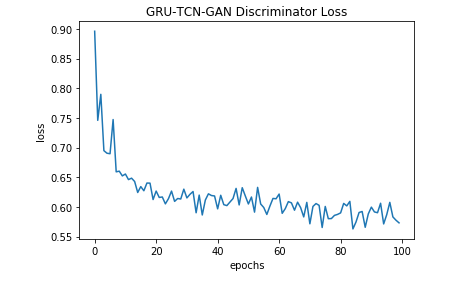}
\caption{Discriminator training loss convergence}
\label{1vsall}
\end{figure}

\section{Results}

To evaluate the quality of the enhanced speech we computed   Perceptual evaluation of speech quality (PESQ) as the major performance metric \cite{rix2001perceptual} for test set speech data and corresponding enhanced speech outputted by the models when the test set data was given as input. We observed that the PESQ value was higher for enhanced speech output compared to that of the test set speech data as shown in Table 1 indicating the enhanced speech output was of better quality than the corresponding test set speech data. 

Since PESQ calculation involve the use of a clean audio signal as reference we computed PESQ values only for five subjects data from test set as only five subjects were common in test set and training set EEG data collection experiments, hence we had a clean reference speech signal only for these five common subjects from the training data set. The average PESQ values for all the test, corresponding enhanced utterances of the five subjects are shown in Table 1. We also observed that our EEG based speech enhancement regression and GAN models outperformed the baseline log MMSE model in terms of PESQ value as seen from Table 1 excpet for TCN regression model where it's test time average PESQ value was similar to the log MMSE model test time PESQ value. The EEG features are not used with log MMSE model, the log MMSE model takes only noisy speech as input and outputs enhanced speech. 
We observed that GRU regression model demonstrated highest PESQ value during test time even though our initial hypothesis was GAN model should demonstrate best results. It shows the difficulty of training GAN models. 

However we computed one more metric namely signal to noise ratio (SNR) for all the test set speech data for the eight subjects and for the enhanced speech outputted by the GRU regression model for all the eight subjects test data input. There are multiple definitions of computing SNR in literature, in our case we computed SNR as ratio of mean to standard deviation of the speech signal. We observed an average SNR value of -8.42e-07 for the test set speech data and average SNR of \textbf{2.62e-06} for enhanced speech outputted by GRU regression model. We can observe that the enhanced speech outputted by the model had higher SNR value compared to the test set data, indicating the enhanced speech outputted was of better quality than the test set speech data. 

We also tried computing another performance metric namely Short Term Objective Intelligibility (STOI) \cite{taal2011algorithm} for the five subjects and we observed average STOI value of 0.020 for test set speech data, 0.0201 with log MMSE model and a highest value of \textbf{0.022} with GRU regression model. The higher STOI value indicates better speech enhancement quality. 

Our overall results demonstrate that EEG features can be used to clean speech recorded in presence of background noise.

\begin{table}[!ht]
\centering
\begin{tabular}{|l|l|l|}
\hline
\textbf{Model}                                                    & \textbf{\begin{tabular}[c]{@{}l@{}}Test Set\\ avg\\ PESQ\end{tabular}} & \textbf{\begin{tabular}[c]{@{}l@{}}Enhanced\\ Output\\ avg\\ PESQ\end{tabular}} \\ \hline
\begin{tabular}[c]{@{}l@{}}Log \\ MMSE\end{tabular}               & 2.4                                                                    & 2.48                                                                            \\ \hline
\textbf{\begin{tabular}[c]{@{}l@{}}GRU\\ Regression\end{tabular}} & 2.4                                                                    & \textbf{2.59}                                                                   \\ \hline
\begin{tabular}[c]{@{}l@{}}TCN\\ Regression\end{tabular}          & 2.4                                                                    & 2.48                                                                            \\ \hline
\begin{tabular}[c]{@{}l@{}}GRU-TCN\\ Regression\end{tabular}      & 2.4                                                                    & 2.52                                                                            \\ \hline
GAN                                                               & 2.4                                                                    & 2.50                                                                            \\ \hline
\end{tabular}
\caption{Speech Enhancement Results}
\end{table}

\section{Conclusion and Future work}

 In this paper we demonstrated cleaning of noisy spoken speech using EEG features recorded in parallel with spoken speech. We make use of state-of-the-art deep learning models like GAN, GRU, TCN regression and EEG signal processing principles to derive our results. To the best of our knowledge this is the first time a spoken speech enhancement using EEG features is demonstrated using deep learning models.

\section{Acknowledgement} 
We would like to thank Kerry Loader and Rezwanul Kabir from Dell, Austin, TX for donating us the GPU to train the models used in this work.

\bibliographystyle{IEEEtran}

\bibliography{refs}
\end{document}